# How we Learn Concepts: A Review of Relevant Advances Since 2010 and Its Inspirations for Teaching

Zhong Wang

(Institute of Clinical Education, Beijing Doers Education Consulting Co., Ltd)

Abstract: This article reviews the psychological and neuroscience achievements in concept learning since 2010 from the perspectives of individual learning and social learning, and discusses several issues related to concept learning, including the assistance of machine learning about concept learning. 1. In terms of individual learning, current evidences shown that the brain tends to process concrete concepts through typical features (shared features); And abstract concepts, semantic processing is the most important cognitive way. 2. In terms of social learning, Interpersonal Neuro Synchronization (INS) is considered the main indicator of efficient knowledge transfer (such as teaching activities between teachers and students), but this phenomenon only broadens the channels for concept sources and does not change the basic mode of individual concept learning. Ultimately, this article argues that the way the human brain processes concepts depends on concept's own characteristics, so there are no "better" strategies in teaching, only more "suitable" strategies.

"How do people learn concepts?" This question is a persistent core topic for education. Because ultimately, any educational idea must be implemented through teaching and learning, and most of the teaching and learning content can be transformed into the learning of concepts, so its importance goes without saying. This article summarizes and reviews some research achievements in psychology and neuroscience since 2010, and attempts to summarize and characterize the latest progress on this issue. At the end of this article, in the discussion section, the author will propose the significance of these developments for teaching improvement.

But before begin the following text, we must first clarify one question: what is a concept? The concept referred to in psychology is not entirely the same as the definition in education. This is highlighted in the existence the term of "core concepts" (also known as "key concept" or "big ideas") in discipline teaching [1]. Hurd believes that the core concept is the conceptual knowledges located at the center of the discipline and which is the backbone of the discipline structure. Obviously, the definition of core concepts cannot be separated from the knowledge structure of the discipline [1]. However, the definition of psychology does not consider this point, and the term "concept" generally refers to things related to classification and judgment. For example, when you see the image of an eagle, you may quickly think of the concept of "bird", but when faced with the question of whether a penguin is considered a bird, experiments have found that most people may hesitate for a moment [2]. Here, the concept of "bird" clearly belongs to a category, and you need to make a judgment that "penguins count as birds" or "penguins do not count as birds". Therefore, the "concept" in psychology generally correspond to the "knowledge point" or "basic fact" in education.

## 1. Individual learning perspective

When a person is quietly reading at their desk, their brain is learning. Focusing on human learning from an individual perspective is an important topic in cognitive science, and the concept learning discussed in this section is all from an individual perspective. The core question from this perspective is: How do we learn/distinguish concepts?

### 1.1 Concrete concepts

Concrete concepts refer to concrete things that are easily accessible in reality, such as "kitten" or "desk". For a long time, the classical view of concepts has been the mainstream of concept learning. This viewpoint holds that people will distinguish whether a thing belongs to a certain concept by analyzing all its features. For example, the concept of "bachelor" must include four elements simultaneously: human, male, adult, unmarried, and all of them are indispensable. However, in 1973, Rips et al. found in their experiment that people's reaction speed to the sentence "robin (Turdidae) is a bird" was significantly faster than that of "turkey is a bird" [3], but in fact, both are birds. This result means that when faced with specific concepts, people are likely not to carefully analyze all the features of these concepts, but rather prefer to find examples with more typical characteristic -robins are obviously more bird like and typical than turkeys -to establish concepts.

Establishing concepts through typical characteristics is not new, but the key issue lies in the underlying mechanisms. This is the primary issue in learning specific concepts. At present, the mainstream explanation for this issue is based on memory encoding theory. Specifically, when learning a concept, the brain encodes it and puts it into the long-term memory; When the encoding is activated by a new event, the brain will retrieve long-term memory based on the activated encoding, which leads to the current event being classified into this concept. Traditional cognitive science believes that this is closely related to the hippocampus [3], which is further supported by recent neuroscience evidence [4]. A result based on machine neural modeling showed that there was a strong linkage and coupling between the hippocampus and the ventromedial prefrontal cortex (vmPFC) during concept recognition. The vmPFC plays an important role in attention regulation [5]. Therefore, the role of typical characteristics for concept discrimination and learning can be understood as the exemplars which with typical characteristics can more attract people's attention, thus making it easier to activate the decoding process of concepts (i.e. the inverse process of encoding), so thus activate the concepts themselves.

Another study revealed that semantics is also involved in this process. The image naming experiment conducted on 78 patients with semantic dementia (SD) in one study showed that the level of conceptual typicality was highly correlated with the completion of tasks by SD patients - the worse the typicality, the worse the results [6]. SD is an obstructive disease in understanding language meanings, believed to be associated with anterior temporal lobe (ATL) atrophy [7]. To verify this result, the researcher conducted the same experiment on 16 normal participants, and conducted transcranial magnetic stimulation interference on their ATL area, obtaining similar results as the previous experiment. This result demonstrates the influence of typicality on the concept of ATL processing [6]. The reason why this result should be worth noting is that in semantic understanding models, the ATL region of the human brain is responsible for the task of conceptual feature splitting. Therefore, this research result showed that typicality characteristics influences human recognition and extraction of conceptual features (such as apples being able to represent sweet, juicy and other features, while greasiness is a distinct feature of avocado) [6].

Although the above achievements roughly outline the brain mechanisms for processing typical characteristics in concept learning, they leave behind a crucial question: what exactly is typical? This question is crucial because it means that teachers need to know which teaching materials are most suitable for concrete concept learning. Unfortunately, the term "typical" is clearly too subjective, and what is typical in everyone's mind may be different. A study published in NeuroImage in 2016 explored this point: researchers recruited 40 participants to complete a task of typical image arrangement. Participants receive randomly images, which may be within the same category or across categories. They need to arrange these images from the most typical to the least typical. During this process, researchers performed functional magnetic resonance imaging (fMRI) scans of their brains. The result showed the meanings of two aspects included in "typicality": the first is the number of shared features. This study suggests that highly typical examples share more features with other examples, while atypical examples contain fewer shared features [8]. For example, apples contain shared features that are more common in fruits (sweet, juicy), while avocado's features (greasy) are difficult to trace in other fruits. The second is the power to distinguishing categories. This study found that categories defined with typical examples are more discriminative than those defined with atypical examples [8]. Taking apples and avocados as examples again, if students use apples to learn the concept of "fruit", they will easily grasp this concept; But if the initial contact is avocado, they will quickly become confused about "fruit" because the characteristics of greasiness is rare in fruits, but is quite common in other foods such as meat. In other words, the characteristic of being greasy brings fruits and non-fruits closer together, blurring the conceptual boundaries. So, "typical" contains two layers of meaning: numbers of share features, and the power of distinguishing categories.

There is still a persistent question about learning concrete concepts: does it require further processing of typical features to establish a concrete concept? In cognitive science, this represents two theories: the view of exemplar and the view of prototype. The former believes that people can construct concepts from examples. For example, Chihuahua, Corgi and Great Dane belong to dogs. The brain will remember these examples and establish the concept of dogs respectively. When encountering a new dog, the brain will retrieve and match these remembered examples one by one, so as to find similar examples [3] [5]. But obviously, this theory leaves some questions, such as how to discriminate a hybrid dog? Because hybridization blurs the boundaries of the samples. Additionally, there is a significant difference in body size between Chihuahuas and Great Danes, why can the brain still easily discover that they are both actually dogs? The prototype view provides an explanation: humans cannot directly establish concepts from examples, but abstract similarities between many typical examples, and then establish a common abstract image. For example, although Chihuahuas and Great Danes have significant differences, they both share the same characteristics as dogs (such as carnivorous, landing on all four limbs), so you would abstract the common features of the two to form an abstract psychological representation. The representation of the abstract dog may not exist in reality, but it can be easily generalized to various specific examples. This abstract process is called category centered tendency [8] or "family similarity structure of concepts" [3].

There has always been controversy over which of these two theories is correct. The results of a fMRI combined with whole brain multivariate modeling (2013) showed that the experimental data tended to be exemplar oriented [9]. While, another similar experiment (2018) showed the opposite result, with 3/4 of the participants exhibiting a clear prototype bias [10]. The controversy over the two is still ongoing, but the prototype view seems to have an advantage. As shown in one of the results in reference [8] of this article, participants showed a clear category center tendency when processing typical examples [8]. Scientists even observed prototype tendencies on two young jackdaws [11]. The most intriguing is a preprint from 2020, which found that prototype and exemplar processing coexisted during the training phase of the task, but unexpectedly disappeared during the task testing phase, resulting in only a prototype bias -even the researchers were surprised by this result [12]. They provided several guesses, one of which was likely a transformation from exemplar representation to prototype representation during the training process. In other words, the participants quietly processed and generalized the examples during the training process, abstracting the commonalities within the examples. The reason why this is intriguing is that the human brain is a living thing, and we try to distinguish the two through experiments, but the brain quietly establishes connections and transformation relationships between them.

## 1.2 Abstract concepts

### 1.2.1 The abstract concepts that can be concretized (CAC)

Similar to concrete concepts, the memory encoding theory mentioned earlier is also a fundamental mechanism for processing abstract concepts [13]. But unlike concrete concepts, abstract concepts are more complex because the first question is: what concept can be considered abstract? Here, there are differences between children and adults, and some concepts that may not seem difficult for adults (such as "ecosystems") are elusive to children due to the lack of prototypes or typical examples. According to Piaget's theory of cognitive development stages, children tend to prefer specific tangible objects before the age of 12 [2]. So, things with blurred boundaries or difficult for children to see or touch are difficult to understand for them.

For this type of concepts (CAC), children exhibit two different tendencies. One of them is to concretize these concepts, that is, to find an example as the prototype of such concepts, so as to transform the task of understanding abstract concepts into cognitive concrete concepts. A study (2023) asked over 2000 Iranian children to draw the image of Allah in their minds. Iran is an Islamic country, and the doctrine stipulates that Allah is intangible, thus prohibiting the worship of any tangible thing. But this is undoubtedly too abstract for children, so some concrete images appeared in their drawings. In one of drawings, a standing woman is staring at the carpet in front of her with her eyes, making it easy to see that she is planning to worship;

In another drawing, there was an elder wearing a robe, much like a theologian. Linking Iran's national conditions suggests that he may be a great Ayatollah -religious leaders with high political and religious status in Iran. Although Allah is extremely abstract, children are not stumped. They find a "shadow" that can represent Allah from the details of life, giving this CAC a visual label.

Other experiments have found that semantic processing can greatly affect children's understanding of CAC. A behavioral experiment from China required 68 fifth grade elementary school students to use arrows to mark out the force on all objects in a painting - which drew six wooden blocks with stationary state at different positions in the water. The results showed that in all cases which without buoyancy arrows, the most were floating wooden blocks, while the least were sinking wooden blocks, and the difference between the two had strong statistical significance. The article argues that this is related to the interfere from the word "floating" (in Chinese, "floating" and "buoyancy" have a same root) [16].

There are also some special CAC that may be related to specific brain regions, with a typical exemplar being causal relationships. Causal relationship is a hidden relationship hidden behind changing things, which is difficult to detect and needs to be reflected through examples. An experiment conducted on 14 adults (with an average age of 25 years) using direct current stimulation combined with fMRI and reaction time in the right parietal lobe showed that the perception of causality is based on sensitivity to spatial and temporal attributes, which are closely related to the function of the right parietal lobe [17]. This result is mutually corroborated by another previous study (2005) [18].

1.2.2 The abstract concepts that are difficult to concretize (DAC)

Although the above concepts are abstract, they can still be prototyped or accurately perceived through experiments. But such as "justice" and "happiness", they are not only abstract, but everyone's definition of them is different. This type of concepts is obviously more difficult to perceive. Therefore, in the following text, we refer to it as "the abstract concepts that are difficult to concretize" (DAC).

An interesting study published in Nature Communication in 2018 provides a unique perspective to explore them. This study recruited 12 congenital blind individuals and 14 normal individuals. It is worth noting that the condition of "congenital blindness" means that they have never seen the appearance of this world since birth. In other words, concepts like "rainbow" that are very concrete to normal people are completely abstract to them, and they cannot even prototype these concepts because they have not even seen the prototype. It was found that when understanding abstract concepts such as morality and justice, the brain regions activated by congenital blind individuals are no different from those of normal individuals. The key to the problem lies in the concepts (such as rainbow and red) that are concrete to normal people. Experiments have shown that blind and normal individuals activate different brain regions - the anterior temporal lobe, which is precisely the activation area of abstract concepts [19]. This seems to reflect a compensatory mechanism of cognitive processing: because congenital blind people cannot learn concepts like "rainbows" through their senses, they can only understand them through language descriptions. This indicates that the processing of concepts in the brain is largely limited by their presentation and acquisition methods [19]. For abstract concepts (especially DAC), semantic processing is one of the most important processing methods.

And how does semantics specifically process DAC? Another study by the research group above (2018) explored the specific organizational principles of abstract concepts. 68 college students were invited to participate in behavioral evaluation of semantic features and subjective semantic distance, while being scanned using fMRI. Researchers processed the results in two different ways and found that both methods can effectively explain the principles of organizing abstract concepts (especially DAC) - lexical co-occurrence and high-dimensional spatial representation. For the former, researchers analyzed the contextual distance of abstract concepts through word co-occurrence patterns (word to word distance) and found a high correlation with the neural response patterns of human language, indicating that word co-occurrence is one of the important organizational forms for understanding abstract concepts. For the latter, researchers constructed a feature space consisting of 13 dimensions, including sensory motor features, emotions, and valence, to represent the processing of abstract concepts. The results show that this semantic processing is more dispersed, that is, the whole brain processing is used to combine and represent semantic features. Researchers further found that valence is one of the extremely special and critical factors. Here, valence includes factors such as attitude, decision-making, predicting the future, personality, etc. The author explains the role of valence: " The special status of valence may be because valence judgment—that is, whether and to what extent a given word is related to positive or negative feelings—lies at the core of human behaviors, exerting strong influence over a wide range of psychological phenomena, including attitudes, decision making, predicting the future, personality, even perception of spoken words and everyday objects" [20].

There are also two ways to understand abstract concepts that cannot be ignored: contextual understanding and association. Although common sense assumes that every word has a precise meaning and usage, in fact, it is not. For example, "spinach" generally appears in contexts related to ingredients and cooking, but the term "life" is not. It can be used in both life sciences or philosophical and religious contexts. This phenomenon is called semantic diversity. Rogers et al. (2013) found that abstract words have stronger semantic diversity, which means they can be used in many different situations [21]. So, how do we determine the specific meanings of these diverse abstract vocabulary? Researchers believe that it is related to semantic control, that is, in a specific context, top-down semantic processing restricts those independent semantics based on the context. For example, in a medical paper, the semantic control mechanism will suppress the religious and philosophical meanings contained in the term "life", thereby accurately locating the semantics related to biological sciences [22].

The study of a patient with post-stroke systemic aphasia (2010) revealed the role of association in abstract concept cognition. The experimenter loudly said one of the words to the patient and asked them to point out that word from a pile of words. For normal people, this is a simple task that can be accomplished by converting speech information into spelling information. However, for that patient, their conversion ability is compromised, and it is obvious that this method is not feasible. The experimental results showed that the patient made more errors in semantically related words (such as cows, horses, sheep) than unrelated words (such as cows, chairs, trees), indicating that patients were likely completing the task through association, converting speech into semantics and further activating adjacent semantics [23].

## 2. Social learning perspective

As we reach the age of 6-7, an extremely important learning channel is added throughout our lifelong learning process: classroom learning (many people may have started this type of learning by attending kindergarten before the age of 3). Unlike before when we mainly relied on individuals to acquire knowledges from the surrounding environment, the classroom is a group with many people. And more importantly, there is usually a leader in learning - the teacher. Generally speaking, the branch of psychology that studies people's mutual influence is called social psychology. Naturally, the learning that occurs in the classroom also falls under the category of "social learning", as it enables you to acquire knowledge through the influence of people (teaching behavior). The core question about concept learning from the perspective of social learning is: how is knowledge (and concepts) transmitted, especially from teachers to students?

Before delving into this section, it must be noted that social learning does not only occur when you enter the classroom, but in fact, it begins from the moment you are born. For example, most people have already learned to speak their mother tongue before the age of two, but almost no one has learned it through their own reading, but rather from their mother and others around them. The reason why this article discusses this issue in a classroom setting is because, on the one hand, compared to infants babbling in the environment, classroom learning is larger in scale and more systematic, and its impact on human cognitive structure is more profound; On the other hand, a considerable number of experiments on this issue are based on classroom scenarios.

Researchers have long known that human learning outcomes are influenced by the behavior of others. A fMRI study (2005) investigated the completion of psychological rotation tasks under the influence of interference and opinions from others. The results showed that 41% of participants would be influenced by others and make wrong choices. Moreover, fMRI results showed that when participants refused to make conformist choices, their emotion related brain regions (such as the right amygdala) were activated. "This means that resisting conformity creates an emotional burden and automatically leads to psychological consumption for those who insist on independence [2]". In other words, conformity is one of the fundamental characteristics of human cognition.

Others can have an impact on learning, which undoubtedly makes teaching behavior possible. But this is no longer the core issue that current cognitive scientists are concerned about, and researchers' interest has shifted to another question: under what conditions is knowledge transfer efficient? A study based on fNIRS hyper scanning technology in 2018 revealed the significance of brain coupling between teachers and students for efficient teaching. This study recruited 60 undergraduate students for pairwise pairing, including a "teacher" and a "student". At the same time, the researchers divided it into three groups and used teaching methods such as "lecture method", "discussion method", and "video teaching" to teach digital reasoning tasks (finding the arrangement rules of a given sequence). When "teachers and students" completing tasks, use near-infrared spectroscopy hyper scanning technology (fNIRS) to perform brain scans on teacher-student pairs. This study found that: ①the teaching effect is best when the activities of the temporoparietal joint area of teachers are synchronized with those of the anterior temporal lobe area of students; ②Single factor ANCOVA analysis shows that video teaching has a significantly weaker ability to activate synchronization compared to the other two methods, while the teaching methods of teaching and discussion have no significant impact on synchronization; ③ The synchronization between teachers and students does not occur simultaneously, there is a time delay of about 10 seconds. That is, the phenomenon of teachers occurring about 10 seconds before students. Furthermore, the researchers established a neurodynamic model of this phenomenon. Ultimately, the article suggests that in efficient teaching activities (i.e.

synchronization between teachers and students), teachers may make predictions immediately before they ask questions and students answer them [24].

There is currently no unified name for the synchronization phenomenon between teachers and students proposed in this article. This article refers to it as "interpersonal neural synchronization (INS)", and some literatures also refers to it as "interbrain synchronization (IBS)" [25]. The INS phenomenon not only exists in teacher-student interaction, but some evidence suggests that it also exists in student group cooperation. A study published in 2023 confirmed this [26]. The study recruited 60 female college students and divided them into 20 groups of three to learn poetry. 20 small groups were randomly divided into two large groups: independent learning and cooperative learning. The requirements of the two groups are completely consistent, with the only difference being that cooperative learning involves a discussion process that explores the main idea, rhetorical devices, and other key points of poetry, while the independent learning group is not allowed to do so. It was found that the cooperative learning group experienced significant intra group neural synchronization (GNS), and the effect of GNS was stronger when group members reached consensus. Further frame by frame analysis shows that GNS can predict learning outcomes around 156-170 seconds (approximately 2 and a half minutes) after the start of learning. The article believes that the results confirm Piaget's viewpoint that successful cooperative learning is related to the moment when learners reach consensus [27].

The primary question about INS is how it was generated? After all, there is no physical connection between the brains of the participants who generate synchronization, so researchers are very curious about the reasons for this phenomenon. Unfortunately, there is still no convincing explanation for this reason, and the only thing we can confirm is that it is time related. For example, an EEG study exploring pianists playing duets (2023) proved this point. Researchers had a performer in a duet intentionally disrupt the rhythm, and found that the disturbance caused a significant decrease in synchronicity. The article points out that INS is actually based on the brain's interpersonal time alignment [28].

Another potential factor is the predictions generated between participants during the interaction process. As mentioned in reference [24], the INS mechanism means that during teacher-student interaction, teachers continuously predict student reactions and adjust their expression strategies. Another more representative evidence is music, where many pieces of music have rhythms that can be predicted. A study on guitar duets (2021) explored the correlation between brain synchronization and interpersonal action coordination, and found that brain synchronization is closely related to the synchronization at the beginning of instrument sound and behavior performance [29]. Another study (2022) directly investigated the impact of rhythm on INS. This study used music with different stresses and frequencies to stimulate 32 female college students and observe their IBS production. It was found that the IBS in the middle frontal cortex (MFC-IBS) is stronger in strong rhythmic music than in weak rhythmic music, resulting in stronger motor coordination. MFC-IBS is widely believed to be related to understanding the psychology and behavior of others in collaboration [30]. Therefore, that study provides crucial evidence for predicting the occurrence of IBS [31].

Meanwhile, this study also provides evidence for another question: how to trigger INS? As mentioned earlier, INS has great practical significance for efficient teaching. But the problem is that we cannot bring fNIRS instruments into the classroom, and teachers cannot constantly adjust their teaching strategies based on the feedback from the instruments. Therefore, it is crucial to determine which behavior can effectively and relatively stably trigger this phenomenon. And the previous research provides an idea for this: it can provide a predictable medium for collaborators, guiding the brains of interactive collaborators to trigger inter brain INS based on this common medium. Other cross-cultural studies have also confirmed the value of medium in activating brain to brain coupling. Due to differences in race, geography, and history, Chinese and Americans have significant cultural differences, making it difficult to inspire INS among them. A study based on video processing (2022) investigated whether synchronization can be activated when watching videos from two cultures. The study recruited 31 pairs of participants, each containing a Chinese and an American. The experiment involves a 20 minute exchange between two individuals in the group, followed by a 5-minute video depicting the lives of high school students from China and the United States, with a clear cultural imprint from both countries. The final results confirm that the presence of cultural others can effectively activate mediators and activate brain coupling between participants (such as mPFC in the ventromedial prefrontal cortex, rTPJ in the right temporal parietal junction, and rMTC in the right middle temporal cortex). For classroom teaching, in addition to the teaching content (music) mentioned in reference [25] as a medium, some simple interactions can also enhance the coupling between teachers and students, such as simple body synchronization before class or nonverbal information interaction between teachers and students during the course [33].

However, the above achievements inevitably raise a causal question: whether behavioral synchronization causes brain synchronization, or brain synchronization causes behavioral synchronization? The above research viewpoint clearly supports the former, while another study proves that the latter also exists. This study recruited 30 couples in love, due to being in a passionate relationship, easily activated their brain synchronization. Transcranial direct current stimulation (tDCS) was used to interfere with their specific brain regions while they completed their common tasks (the true stimulation group stimulated the right anterior temporal lobe rATL, the sham stimulation group did not, and the control group stimulated the occipital lobe). Due to the presence of tDCS, the INS between couples has been disrupted. Changing interpersonal INS through external non-invasive stimuli is a classic research paradigm, and Giacomo Novembre et al. were one of the earliest explorers of this paradigm. They found that transcranial alternating current stimulation (tACS) can significantly enhance INS among collaborators performing right-handed tapping tasks [34]. And this tDCS study has two main findings: ① there was no significant change in behavior between couples, but their level of emotional empathy significantly decreased; ② Mediation effect analysis shows that nonverbal behavior mediates the association between decreased INS and decreased empathy. This study suggests that changes in INS can affect implicit cognition by altering nonverbal behavior [35]. From the above evidence, it appears that there is a bidirectional interaction between human behavior and the coupling of the brain: on the one hand, some behaviors can trigger INS; On the other hand, changes in INS can also trigger changes in behavior (such as nonverbal behavior).

Since this article focuses on how people learn concepts, we are obviously more concerned about how the efficient transmission of knowledge triggered by INS affects individual concept learning, i.e., how it embeds into the individual concept learning mode mentioned earlier? Unfortunately, due to limitations in instruments and measurement techniques, this issue cannot be fully understood at present. Due to the complete intolerance of artifacts and high cost, fMRI is difficult to adapt to real-world research on teaching and learning, while fNIRS and EEG are more suitable in this regard [36]. Among them, fNIRS has developed the most rapidly because compared to EEG, fNIRS can understand the spatial changes of blood oxygen in brain activity through hemodynamic responses, making the evidence more intuitive and reliable [37]. However, fNIRS can only monitor changes in blood oxygen levels in the superficial cortex [37], but deep brain regions such as the hippocampus and amygdala play a crucial role in concept learning [38], and fNIRS is powerless in these areas. Therefore, little is known about the impact of INS on concept learning.

However, based on existing research findings, there is no evidence to suggest that INS triggers a completely different concept learning mode from individual concept learning. As stated in reference [24] of this article, although teaching outcomes are related to the activation of INS, the correlation with teaching methods or styles is not significant with outcomes [24]. In other words, although social learning enriches the channels for concept learning, it has not changed the basic mode of concept learning.

At the same time, by the single brain scanning, researchers have also discovered another interesting phenomenon of teaching leading to concept learning: if students already have an incorrect prior understanding of the concepts being taught, what impact will teaching have on them? A study (2019) combining fNIRS with eye tracking to assess students' mastery of scientific concepts revealed that knowledge transfer affects individual concept learning by regulating cognitive inhibition. This study presents the concept of the causes of the four seasons to participants, in order to explore the differences between those who master daily concepts and those who have correct concepts. The Earth's orbit around the Sun is an ellipse with perihelion and aphelion, but unfortunately, for the Northern Hemisphere, summer happens to be at aphelion and winter at perihelion. This phenomenon contradicts intuition and is a preconception that is prone to errors for students. This study found that although students were taught that preconceptions were incorrect, eye skip data showed that this misconception did not disappear, but was suppressed due to cognitive inhibition (fNIRS showed significant activation of the brain area responsible for inhibition in the right prefrontal lobe) [39]. This study reveals what the competition mechanism in the brain is if there is a significant difference between old and new knowledge.

## 3. Some other issues

3.1 How to distinguish easily confused concepts?

The classification of concepts is not always easy to achieve, and often it is easy to make mistakes. It is not difficult to understand abstract concepts being misclassified because, as mentioned earlier, many of them require implicit processing. However, there is also a confusion issue with concrete concepts that have always been regarded as exact, which has sparked the interest of researchers.

The confusion of specific concepts can be roughly divided into two types, as described in reference [6] of this article regarding the difference between apples and avocados [6]. Avocado, due to its limited shared features, is not easily classified as a "fruit", which clearly belongs to the issue of the number and relevance of shared features. This phenomenon is very common in children, and due to a lack of knowledge, they are prone to categorizing "whales" as "fish". Due to convergent evolution, whales resemble fish very much. And for this easily confused concept caused by shared features, it is clearly the issue of the anterior temporal lobe region [6].

But another type of confusion is completely different: how to distinguish between strawberries and raspberries? These two have similar appearance and belong to the Rosaceae shrub family, with their edible parts being their berries. Obviously, they are close relatives with only minor differences. For this type of easily confused concept, evidence points to the granularity of features. For example, a study (2014) pointed out the contribution of the perinasal cortex (PC) to this. This study used fMRI to scan the brains of participants who completed the task of naming confusing concept images, and the results of representational similarity analysis (RSA) showed that PCs distinguished these concepts through more fine-grained processing. The article further found that increasing the degree of confusion of concepts will stimulate the activity of PC to a greater extent [40]. Another study on Alzheimer's disease (AD) patients (2012) confirmed this conclusion from the perspective of mental illness. The study found that thinning of the perinasal cortex caused by AD significantly disrupted patients' ability to distinguish easily confused concepts [41]. In short, different types of easily confused concepts will stimulate different brain regions for discrimination processing.

3.2 Is the concept learning mode of different disciplines completely different?

This article focuses on the general laws related to concept learning, so they should be applicable to the learning of the vast majority of disciplines. However, there are significant differences between different disciplines, so it is evident that the specific patterns of conceptual processing in different disciplines are different.

At present, the main evidence for this tends to be related to the similarity of disciplinary content. For example, an article in 2022 explored the similarities and differences in learning mathematics, physics, and chemistry. The article believes that overall, due to the high degree of integration and similarity among these three disciplines, their neural processing methods have a high degree of similarity. For example, they all have spatial processing content (such as spatial transformations in geometry or chemical structures similar to benzene rings in chemistry), so their visual spatial network (parietal lobules and frontal gyrus) processing is very similar. Similarly, all three have a large number of abstract concepts, so they also activate similar semantic processing areas (such as the middle temporal gyrus, inferior frontal gyrus, angular gyrus, dorsomedial prefrontal cortex, etc.) [42].

However, these three are different after all, so there are also some differences in processing. Mathematics is relatively unique, as it is more abstract, resulting in a higher degree of activation of the visual spatial network during mathematical learning. This may indicate that the brain is trying to imagine what those abstract mathematical formulas look like. On the other hand, arithmetic is clearly also a part of mathematics. However, this study found that arithmetic almost did not activate the semantic regions of the subjects, but there was a clear involvement of semantics in completing physical and chemical learning tasks.

This result is also supported by another study from another perspective: this study (2016) used fMRI to compare the differences in activation networks between mathematicians and humanists when dealing with mathematical and non mathematical tasks. Due to the significant differences between mathematics and humanities, there should be significant differences in the processing of mathematics between these two types of scholars. The fact also proves that mathematicians have a significantly stronger level of frontal parietal lobe activation when completing mathematical tasks than humanities [43].

3.3 How can machine learning be applied to concept learning research?

In recent years, the use of machine learning for auxiliary research in cognitive research has become increasingly popular, and some scholars have even voluntarily abandoned traditional psychological experiments and embraced machine learning (ML) [44]. Although this viewpoint may seem too radical, it is not difficult to see the significant impact of ML on the traditional cognitive research paradigm. Since this article is not specifically focused on this issue, readers can refer to Zhihua Zhou's book "Machine Learning" [45] or Mitchell T's book "Machine Learning" [46] for basic introductions to ML and artificial intelligence. Regarding the content of neural networks and deep learning, it is particularly recommended to read Juergen Schmidhuber's article "Deep Learning in Neural Networks: An Overview" [47].

In terms of the research paradigm itself, in the field of application, ML can help researchers better partition concepts through pre-training data. For example, many patients may encounter misdiagnosis problems. Especially for some rare diseases, the probability of misdiagnosis is even higher. Some rare diseases share symptoms with many common diseases (such as idiopathic pulmonary hypertension and coronary heart disease, both of which can trigger angina). In other words, the conceptual boundary between rare diseases and common diseases is not clear in the minds of initial diagnosis doctors. ML, on the other hand, can use pre trained massive data to identify the most likely cause and provide doctors with specific questions or diagnostic recommendations to assist them [48].

Secondly, simulating human cognitive activities through ML is one of the most important and emerging research paradigms in current cognitive science. As mentioned in reference [5] of this article, traditional cognition focuses on dissociating different brain regions according to their functions, such as the ventromedial prefrontal cortex responsible for concept generalization, and the anterior temporal lobe responsible for semantic processing and understanding of concepts [5]. However, the processing of concepts is a complex process that often requires the cooperation of different functional brain regions to complete. Therefore, the collaborative mechanism of multiple brain regions is one of the most concerned topics in current cognitive science, and the development of ML provides crucial support for it [5].

We take the SUSTAIN model as an example, which clusters perceptual information according to attention weights, distinguishing it into different feature dimensions (such as color and shape), and forming some clusters separately. These clusters are matched with specific categories of storage and ultimately achieve the goal of classifying concepts. Therefore, SUSTAIN is particularly suitable for simulating or validating the classification processes for concrete concepts. For example, Michael L. Mack et al. (2016) used this model to verify the dynamic changes of hippocampus when people learn new concepts. The study recruited 23 volunteers who were required to classify insects according to different rules (for the previous classification, the new classification rules were equivalent to learning a new concept) and their brain activity was scanned using fMRI. Then, the researchers applied the same task to SUSTAIN to examine whether it could reproduce the performance of participants, and evaluated the similarity between the two using Spearman correlation and randomization tests. The study found that when tasks changed, the hippocampus could organize knowledge in new ways, generating new concepts, and SUSTAIN validated this result [49].

In addition to concrete concepts, ML has also been proven to be useful for studying the learning of abstract knowledge and implicit processes. For a child, what constitutes "adult" is an abstract concept, as it not only implies a tall body, but also behavior model with completely different purposes from that of a child. Therefore, we often see children attempting to imitate adult behavior through observation and learning. A study (2023) fitted this imitative behavior by using ML. This study investigated whether children's behavior imitation towards adults includes reasoning about adult intentions. For this, they constructed two computational models, one assuming the existence of this implicit process, and the other assuming the absence. The results show that the former has a better explanatory effect on young children's behavior. This result has also been validated by the fNIRS results [50].

Besides verifying experimental results, the most commonly thought of use of ML is for mining and analyzing experimental data, which is not difficult to understand, as computers have much greater data processing capabilities than humans. However, this raises a natural question: what are the advantages of ML compared to some professional statistical software such as IBM's SPSS? In other words, what is the difference between statistics and ML? Graziella Orrù et al. argue that statistical methods focus on whether the results are "real" effects or caused by noise (p-value testing is usually the most commonly used method for this approach); The ML method treats data rulers as unknowns and focuses on predicting unobserved results or future behavior [51]. In other words, ML is more suitable for discovering patterns from data, while statistical methods are more suitable for verifying the discovered patterns.

## 4. Inspiration for Education

The main inspiration for education in this article may be that there are no "better" methods in teaching and learning, only "more suitable" methods.

Both frontline teachers and educational researchers are tirelessly seeking which learning modes are better and more in line with the cognitive laws of students. Proceed to the next step, develop teaching strategies that align with these learning modes in order to promote student learning more efficiently. Taking inquiry based learning (LBL) as an example, inquiry based learning and its teaching model (IT) have been one of the most advocated methods in international science education for decades [52], and have also been endorsed by authoritative and influential educational evaluation projects such as PISA (Program of International Student Assessment) [53]. The result of this advocacy and endorsement is that IT has gradually become the mainstream of frontline teaching (especially science education), and there are even signs of abuse. A survey of junior high school chemistry classes in Chinese Mainland (2020) shows that quite a number of teachers can't tell when IT is more suitable for use, and even mistakenly believe that chemistry teaching should be an IT class based on experiments [54].

However, from the literature reviewed in this article, it can be seen that different types of learning content have completely different cognitive processing patterns in the human brain. If we force all learning content to adopt the same learning method, abuse is inevitable. In fact, a considerable amount of science teaching content is not suitable for using inquiry methods, such as Canxi Cao and other XBGoost based machine learning methods, which have found that reading has great value in improving scientific literacy, and reading is clearly far from traditional inquiry methods [55]. Moreover, there has always been controversy over whether LT is indeed effective. For example, a covariance multivariate analysis of students learning electrical concepts, scientific process skills, and attitudes showed that LT did not show significant advantages compared to the teaching method [56]. Another comparative study (2021) based on PISA2015 data on IT effectiveness in six countries has yielded puzzling conclusions: on the one hand, students who receive high-level teacher guidance have higher scientific literacy; On the other hand, the scientific literacy of received high-level IT students is actually lower. Even researchers lack confidence in the consistency of this result, so at the end of the article, it is mentioned that "in science teaching, we do not advocate one teaching method over other teaching methods" [57]. Feng Jiang et al. more directly pointed out the fundamental problem in current IT: most of the evidence supporting IT comes from pure research environments similar to laboratories, rather than real school learning contexts [58].

Fortunately, many educational researchers have realized the significance of using experiments to find the most suitable teaching methods for students, rather than just following authority or mainstream. For example, He and Oltra Massuet evaluated learners' grammar sensitivity and language production ability through oral imitation task tests to construct English problems with fine grammar errors [59]. For another example, Ayabe et al. studied the value of charts in mathematics teaching and found that including charts that are suitable for problems can help improve student problem-solving performance and reduce cognitive burden in problem-solving activities [60]. In summary, as mentioned earlier, different types of concepts have different cognitive approaches, so educators should explore more suitable teaching methods and strategies based on the teaching content.